\documentstyle [12pt]{article}
\title{Ground-state properties of the Falicov-Kimball model 
with correlated hopping in two dimensions}
\author{Pavol Farka\v sovsk\'y and Nat\'alia Hud\'akov\'a\\
Institute  of  Experimental  Physics,  Slovak   Academy   of
Sciences\\
Watsonova 47, 043 53 Ko\v {s}ice, Slovakia}
\date{}
\begin{document}
\baselineskip=20pt
\maketitle

\begin{abstract}
A new numerical method, recently developed to study ground states 
of the Falicov-Kimball model~(FKM), is used to examine the effects 
of correlated hopping on the ground-state properties of this model
in two dimensions. It is shown that the ground-state 
phase diagram as well as the picture of metal-insulator transitions 
found for the conventional FKM (without correlated hopping) are strongly 
changed when the correlated hopping term is added. The effect of correlated 
hopping is so strong that it can induce the insulator-metal transition, 
even in the strong-coupling limit, where the ground states of the 
conventional FKM are insulating for all $f$-electron densities.

\end{abstract}
\thanks{PACS numbers.:75.10.Lp, 71.27.+a, 71.28.+d, 71.30.+h}

\newpage
\section{Introduction}
Since its introduction in 1969, the FKM~\cite{Falicov}
has become an important standard model for a description of
correlated fermions on a lattice. The model was originally 
proposed to describe metal-insulator transitions and has since
been investigated in connection with a variety of problems such
as binary alloys~\cite{Freericks_Falicov}, the formation of 
ionic crystals~\cite{Gruber/Lebowitz}, and ordering in mixed-valence 
systems~\cite{Ramirez}. It is the latter language we shall use here, 
considering a system of localized $f$ electrons and itinerant $d$ 
electrons coupled via the on-site Coulomb interaction $U$. 
The Hamiltonian of the spinless FKM is 

\begin{equation}
H=\sum_{ij}t_{ij}d^+_id_j+U\sum_if^+_if_id^+_id_i+E_f\sum_if^+_if_i,
\end{equation}
were $f^+_i$, $f_i$ are the creation and annihilation 
operators  for an electron in  the localized state at 
lattice site $i$ with binding energy $E_f$ and $d^+_i$,
$d_i$ are the creation and annihilation operators 
for an electron in the conduction band. The conduction 
band is generated by the hopping 
matrix elements $t_{ij}$, which describe intersite
transitions between the sites $i$ and $j$. 
Usually it is assumed that $t_{ij}=-t$ if $i$ and $j$ 
are nearest neighbors and $t_{ij}=0$ otherwise (the conventional FKM), 
however, in what follows we consider a much more realistic  
type of hopping, so for the moment we leave it to be arbitrary.

Recent theoretical works based on exact numerical and analytical calculations 
showed that the FKM, in spite of its relative simplicity, 
can yield the correct physics for describing of such fundamental 
phenomena as valence-change transitions, metal-insulator transitions,
crystallization, charge ordering, etc. 
For example, it was found that the spinless FKM, in the 
pressure induced case, can describe both types of intermediate-valence
transitions observed experimentally in rare-earth compounds: a discontinuous 
insulator-insulator transition for sufficiently strong 
interactions~\cite{Fark1} and a discontinuous insulator-metal transition 
for weak interactions~\cite{Fark2}. In addition, at nonzero temperatures this 
model is able to provide the qualitative explanation for anomalous large 
values of the specific heat coefficient and for extremely large changes of 
electrical conductivity~\cite{Fark3} found in some intermediate-valence 
compounds (e.g., in SmB$_6$).  Moreover, very recently the spin-one-half 
version of the FKM has been used to describe a  discontinuous 
intermediate-valence  transition (accompanied by a discontinuous 
insulator-metal transition) in SmS~\cite{spin} as well as 
for a description of an anomalous magnetic response of the Yb-based 
valence-fluctuating compounds~\cite{Freericks1}.

On the other hand it should be noted that the model Hamiltonian~(1)
neglects all nonlocal interaction terms, and thus it 
is questionable whether above mentioned results persist also in
more realistic situations when nonlocal interactions will be turned 
on. An important nonlocal interaction term obviously absent
in the conventional FKM is the term of correlated hopping,
in which the $d$-electron hopping amplitudes $t_{ij}$ between neighboring 
lattice sites $i$ and $j$ depend explicitly on the occupancy $(f^+_if_i)$ 
of the $f$-electron orbitals. To examine effects of this term on 
ground-state properties of the two-dimensional FKM we choose the 
following form for the nearest-neighbor matrix elements

\begin{equation}
\tilde{t}_{ij}=t_{ij}+t'_{ij}(f^+_if_i+f^+_jf_j),
\end{equation}
which represent a much more realistic type of electron hopping 
than the conventional hopping.

Thus the spinless FKM in which the effects of correlated hopping are
included can be written as
\begin{equation}
H=\sum_{<ij>}t_{ij}d^+_id_j+\sum_{<ij>}t'_{ij}(f^+_if_i+f^+_jf_j)d^+_id_j+
U\sum_if^+_if_id^+_id_i+E_f\sum_if^+_if_i.
\end{equation}
The first term of (3) is the kinetic energy corresponding to
quantum mechanical hopping of the itinerant $d$-electrons
between the nearest-neighbor sites $i$ and $j$. 
The second term is just the correlated hopping term discussed above.
The third term is the on-site Coulomb 
interaction between the $d$-band electrons with density
$n_d=\frac{1}{L}\sum_id^+_id_i$ and the localized
$f$-electrons with density $n_f=\frac{1}{L}\sum_if^+_if_i$, where $L$ 
is the number of lattice sites. 
The last term stands for the localized $f$ electrons whose sharp 
energy level is $E_f$.

Since in this spinless version of the FKM
without hybridization  the $f$-electron occupation
number $f^+_if_i$ of each site $i$ commutes with
the Hamiltonian (3), the $f$-electron occupation number
is a good quantum number, taking only two values: $w_i=1$
or 0, according to whether or not the site $i$ is occupied
by the localized $f$ electron. Therefore the Hamiltonian (3) can 
be written as

\begin{equation}
H=\sum_{<ij>}h_{ij}(w)d^+_id_j+E_f\sum_iw_i,
\end{equation}
where $h_{ij}(w)=\tilde{t}_{ij}(w)+Uw_i\delta_{ij}$ and

\begin{equation}
\tilde{t}_{ij}(w)=t_{ij}+t'_{ij}(w_i+w_j).
\end{equation}

Thus for a given $f$-electron configuration
$w=\{w_1,w_2 \dots w_L\}$ the Hamiltonian (4)
is the second-quantized version of the single-particle
Hamiltonian $h(w)$, so the investigation of
the model (4) is reduced to the investigation of the
spectrum of $h$ for different configurations of $f$ electrons.
Since the $d$ electrons do not interact among themselves, the
numerical calculations should precede directly in the following steps.
(i) Having $w=\{w_1,w_2 \dots w_L\}$, $U$, $E_f$ and the nearest-neighbor
hopping amplitudes $t$ and $t'$ fixed, (in the following $t=-1$ and all 
energies are measured in units of $t$) find all eigenvalues 
$\lambda_k$ of $h(w)$. (ii) For a given
$N_f=\sum_iw_i$ determine the ground-state energy
$E(w,U,E_f)=\sum_{k=1}^{N-N_f}\lambda_k+E_fN_f$ of a particular
$f$-electron configuration $w$ by filling in the lowest
$N_d=N-N_f$ one-electron levels
(here we consider only the case $N_f+N_d=L$,
which is the point of the special interest for valence
and metal-insulator transitions caused by promotion of electrons
from localized $f$ orbitals $(f^n \to f^{n-1})$ to the conduction
band states). 
(iii) Find the $w^0$ for which
$E(w,U,E_f)$ has a minimum. Repeating this procedure for different
values of $U,t'$ and $E_f$, one can study directly  
the ground-state phase diagram and  valence 
transitions (a dependence of the $f$-electron occupation number on the 
$f$-level position $E_f$) in the FKM with correlated hopping.

A direct application of this method has been used successfully  
in our previous papers~\cite{Fark1,Fark2} for a description 
of ground-state properties of the one-dimensional FKM model 
without  correlated hopping ($t'=0$). 
It was shown that finite-size effects are negligible for a wide
range of the model parameters (e.g., strong interactions) 
and thus results obtained on relatively small clusters ($L<30$) can be 
satisfactory extrapolated to the thermodynamics limit ($L\to \infty$).
Using this method we have described satisfactory
the strong-coupling phase diagram as well as the picture of valence and
metal-insulator transitions in the one-dimensional spinless 
FKM~\cite{Fark1} with $t'=0$. It was found that for sufficiently 
large $U$ the spinless FKM undergoes only a few discrete
intermediate-valence transitions. These intermediate-valence transitions  
are insulator-insulator transitions, since they are
realized between the insulating ground states corresponding to
the most homogeneous configurations, which are the ground states
in this region~\cite{Lem}. Thus, there are no insulator-metal 
transitions in the 1-d conventional FKM for strong interactions. 
In the next paper~\cite{Fark4} we have shown 
that this picture of valence and metal-insulator transitions is 
dramatically changed if the term of correlated hopping is included. 
One of the most important results found for the one-dimensional
FKM with correlated hopping was that the correlated hopping can
induce the insulator-metal transition, even in the half-filled band
case $n_d+n_f=1$. In this paper we try to show that the same result holds 
also for the two dimensional case. Similar calculations are performed also
away from the half-filled band case with the goal to examine possibilities 
for metal-insulator transitions in the strong-coupling limit.  
Another inspiration for performing these calculations was the recent 
paper of Wojtkiewicz and Lemanski~\cite{Woj_Lem}, where the authors 
studied two-dimensional FKM with correlated hopping using the combination 
of the perturbation expansion (up to the second order) and the method 
of restricted phase diagrams. They found that only a few phases form the 
ground-state phase diagram of the model in the strong coupling limit. 
For example, the ground state of the model for $E_f=0$ is the 
chessboard charge-density-wave~(CDW) phase for all $0<t'<1$.
Here we show that some other configurations, (e.g., the segregated 
configuration) can also be the ground states of the FKM 
at $E_f=0$, thereby the ground-state phase diagram 
as well as the picture of metal-insulator transitions are strongly
changed.

\section{The method}
Since the number of configurations that should be
examined to obtain the ground state energy of the FKM grows exponentially
with the system size, a direct application of the exact-diagonalization
method described above is restricted to clusters up to 30 sites.
In our previous papers we showed that clusters of this size are sufficient 
to suppress finite-size effects in one-dimension~\cite{Fark1,Fark2}, 
however, to obtain trustworthy results on the ground-state energy of the model 
in two dimensions one has to examine much larger clusters ($L \sim 100$).
Unfortunately, the clusters with $L>30$ are beyond the reach of present
day computers within exact diagonalizations, and thus the only way
is to compute the ground-state properties of the model by an
approximate but well controlled method.
Here we use the simple method based on a modification of the
exact-diagonalization procedure described above. 
The method consists of following steps: 
(i) Chose a trial configuration $w=\{w_1,w_2 \dots w_L\}$.
(ii) Having $w$, $U$ and $E_f$ fixed, find
all eigenvalues $\lambda_k$ of $h(w)=T+UW$. (iii) For a given
$N_f=\sum_iw_i$ determine the ground-state energy
$E(w)=\sum_{k=1}^{L-N_f}\lambda_k+E_fN_f$ of a particular
$f$-electron configuration $w$ by filling in the lowest
$N_d=L-N_f$ one-electron levels.
(iv) Generate a new configuration $w'$ by moving a randomly
chosen electron to a new position which is chosen also at random.
(v) Calculate the ground-state energy $E(w')$. If $E(w')<E(w)$
the new configuration is accepted, otherwise $w'$ is rejected.
Then the steps (ii)-(v) are repeated until the convergence
(for given parameters of the model) is reached.
Of course, one can move instead of one electron (in step (iv))
two or more $f$ electrons, thereby the convergence of method can be 
improved. Indeed, tests that we have performed for a wide range of the 
model parameters showed that the latter implementation of method, 
in which $N_0 > 1$ electrons ($N_0$ should be chosen at random) 
are moved to new positions, overcomes better the local minima of 
the ground-state energy. This also improves the accuracy of method. 

This method was first used in our recent paper~\cite{Fark5} to study
the ground-state properties of the one and two-dimensional FKM 
without correlated hopping. 
It was found that on small and intermediate clusters ($L\sim 30$) 
the method is able to reproduce exactly the ground states of the 
conventional FKM, even after relative small number of iterations 
(typically 10000 per site). For such clusters the method is only
rarely stopped at the local minimum. Of course, with increasing $L$
the problem of local minima appears often. Fortunately, it can be 
considerably reduced by more efficient algorithm (one is discussed 
above) or by increasing the number of iterations.
The latter case imposes, however, severe restrictions on the size
of clusters than can be studied with this method ($L\sim 100$, for 
$10^6$ iterations per site).  
To verify the convergence of this method for the two-dimensional 
FKM with correlated hopping we have performed the same 
calculations on the cluster of $4 \times 4$ sites, where ground
states can be obtained also within the exact diagonalization calculations.
Numerical results obtained for a wide range of the model parameters 
($t'=-1,-0.8,\dots 1, U=0,0.1 \dots 10$) shoved that the 
exact ground-states can be again reproduced after $\sim10000$ 
iterations per site.

\section{Results and discussion}

The most interesting question that arises for the FKM with 
correlated hopping is whether the correlated hopping term can change the 
ground-state phase diagram and the picture of valence and metal-insulator 
transitions found for the conventional FKM  ($t=-1$ and $t'=0$). 
The nature of the ground state, its energetic and structural properties, and 
the correlation-induced metal-insulator transitions are subjects of special
interest. For the conventional FKM  these problems are well understood at 
least in the symmetric case ($E_f=0,n_f=n_d=1/2$). In this case
the localized $f$-electrons fill up one of two sublattices of 
the hypercubic lattice (the charge-density-wave state) 
and the corresponding ground state is insulating for all~$U>0$. 
Thus, for the finite interaction strength there is no correlation-induced 
metal-insulator transition in the symmetric case. 

One can expect, on the base of simple arguments, that the ground-state 
phase diagram of the FKM with correlated hopping will be  fully different 
from one discussed above for the conventional FKM. 
Indeed, the following selection of hopping matrix amplitudes 
$t=-1$ and $t'>0$ may favor the segregated configuration 
since the itinerant $d$ electrons have the lower 
kinetic energy in this state. This mechanism could lead, for example,  to the
instability of the CDW state that is the ground state for 
$t'=0$, and thereby to a metal-insulator transition, even in the symmetric 
case. To examine possibilities for such a transition in two dimensions 
we have performed an exhaustive study of the model on $6 \times 6$ and 
$8 \times 8$ clusters (with periodic boundary conditions) 
for a wide range of parameters  $t'$ and $U$. 
The results of numerical calculations are summarized in 
Fig.~$1$ in the form of the $t'$-$U$ phase diagram. In addition to the 
CDW state $w_{1}$ that is the ground state at 
$t'=0$ for all nonzero $U$ we found two new phases that can be the ground
states of the model, and namely, the configuration with alternating lines
of occupied and unoccupied sites $w_2$ and the segregated configuration 
$w_3$ (see Fig.~2). Thus at nonzero $t'$ the CDW state $w_{1}$ 
becomes unstable against the transition to $w_2$ and $w_3$. 
The transition from $w_{1}$ to $w_2$, as well as from $w_{1}$ to $w_3$ 
is the insulator-metal transition since the configuration $w_1$
has the finite gap ($\sim U$) at the Fermi energy~\cite{note} for all
nonzero values of $U$, while both $w_{2}$ and $w_3$ are metallic in the
corresponding regions of stability.
Thus we arrive to the very important conclusion, and namely, that 
the correlated hopping term can induce the insulator-metal transition,  even 
in the half-filled band case. Another important result, confirming  
the crucial role of the correlated hopping term can be seen from Fig.~1, 
where the comprehensive phase diagram of the two-dimensional FKM with 
correlated hopping is presented.  It is seen that the correlated hopping 
can destroy the CDW state, even at large values of the Coulomb 
interaction~($U\sim 7$). This is an unexpected result since recent results 
of Wojtkiewicz and Lemanski~\cite{Woj_Lem} based on the combination of 
the perturbation expansion (up to the second order) and the method of 
restricted phase diagrams predicted that the ground state of the model 
at $E_f=0$ and $U$ large is the CDW state for all values of $0< t'< 1$. 
This discrepancy is probably due to the fact that the authors examined 
(as possible ground states) only a restricted set of configurations
(consisting of all periodic configurations having elementary cells up 
to 12 sites), and the segregated configuration
(that should be the ground state in this region) does not belong 
to this set. Another possible explanation of this discrepancy is that 
the second-order perturbation expansion used by authors is insufficient
to describe correctly the ground-state properties of the model 
in this region ($U\sim 7$).  
  
The fact that the correlated hopping can induce metal-insulator 
transitions indicates that the picture of valence and metal-insulator 
transitions found in our previous papers within the conventional 
FKM~\cite{Fark1,Fark2,Fark5} should be dramatically 
changed if finite values of $t'$ will be considered. 
The largest changes are expected in the strong coupling limit 
($U>4$), where all ground states of the conventional FKM are insulating 
for both 1-d and 2-d case~\cite{Fark1,Lem,Fark5}, 
while the numerical results obtained for
nonzero $t'$ show on the existence of the metallic phase,
at least for $n_f=1/2$. We suppose that this important result is not 
restricted to the half-filled band case only, but persists also
for $f$-electron densities away from this point.   
To verify this conjecture we have performed an exhaustive study of the model 
for $n_f=1/4$ on $6\times6$, $8\times8$ and $12\times12$ clusters.
Our numerical calculations showed that the phase diagram of the model
at $n_f=1/4$ is separated into two distinct regions. In the first
region ($U<2$) the phase diagram has a complex structure with the
ground state apparently changing point by point at every value 
of the correlated hopping amplitude $t'$ for fixed interaction 
strength. Unfortunately, the structure of the phase diagram 
in this region strongly depends on the size of cluster and thus
we were not able to extrapolate  satisfactory these results to the 
thermodynamic limit $L\to \infty$. Contrary to this case, the phase diagram 
exhibits a very simple structure~(see inset in Fig.~1) in the opposite 
limit ($U>2$). In this region only two configurations are the ground states 
of the FKM with correlated hopping, and namely the segregated configuration 
and the configuration $w_4$ (see Fig.~2) that has been proven to be ground 
state of the  conventional FKM for 
large~$U$~(see Ref.~\cite{Fark5,Kennedy,Watson}). 
Since the configuration $w_s$ is metallic and $w_4$ insulating 
we have the correlated hopping induced metal-insulator transitions also
at $n_f=1/4$. The metallic phase is stable up to $U \sim 7$ and this again 
confirms our conjecture that the comprehensive picture of metal-insulator 
transitions in the FKM with correlated hopping will be 
fully different from one found for the conventional FKM, 
especially for $U$ large. 
To complete this picture one has to perform similar calculations
for all $f$-electron densities what is a cumbersome computational task,
even on $8\times 8$ cluster. The work on this subject is currently 
in progress.

In summary, the effects of correlated hopping on the ground-state 
properties of the FKM in two dimensions have been studied. 
It was shown that the ground-state phase diagram as well as the picture 
of metal-insulator transitions found for the conventional FKM are strongly 
changed when the correlated hopping term is added. The effect of correlated 
hopping is so strong that it can induce the insulator-metal transition, 
even in the strong-coupling limit, where the ground states of the 
conventional FKM are insulating for all $f$-electron densities.

\vspace{0.5cm}
This work was supported by the Slovak Grant Agency VEGA
under grant No. 2/7021/20. Numerical results were obtained using
computational resources of the Computing Centre of the Slovak 
Academy of Sciences.

\newpage
Figure Captions

\vspace{0.5cm}
Fig.~$1.$ 
$t'$-$U$ phase diagram of the two-dimensional FKM with correlated hopping at 
half-filling ($E_f=0,n_f=n_d=0.5$). Three different phases correspond to the 
CDW state $w_{1}$, the configuration with alternating lines of occupied 
and unoccupied sites ($w_2$), and the segregated configuration $w_3$. 
The inset shows $t'-U$ phase diagram for $n_f=1/4$ and $U>2$.
Two different phases correspond to the segregated configuration 
and the configuration $w_4$ that has been proven to be ground state of the 
conventional FKM for large $U$.

\vspace{0.5cm}
Fig. $2.$
The ground-state configurations of the two dimensional FKM
with correlated hopping for $n_f=1/2$ ($w_1,w_2$ and $w_3$) and
$n_f=1/4$ ($w_4$).  


\begin{thebibliography}{99}
\bibitem{note} Since we consider the case $N_f+N_d=L$,
the Fermi level $E_F$ and the energy gap $\Delta$
of some configuration $w$ are given by $E_F=\lambda_{L-N_f}$ and
$\Delta=\lambda_{L-N_f+1}-\lambda_{L-N_f}$, respectively.
\end{thebibliography}
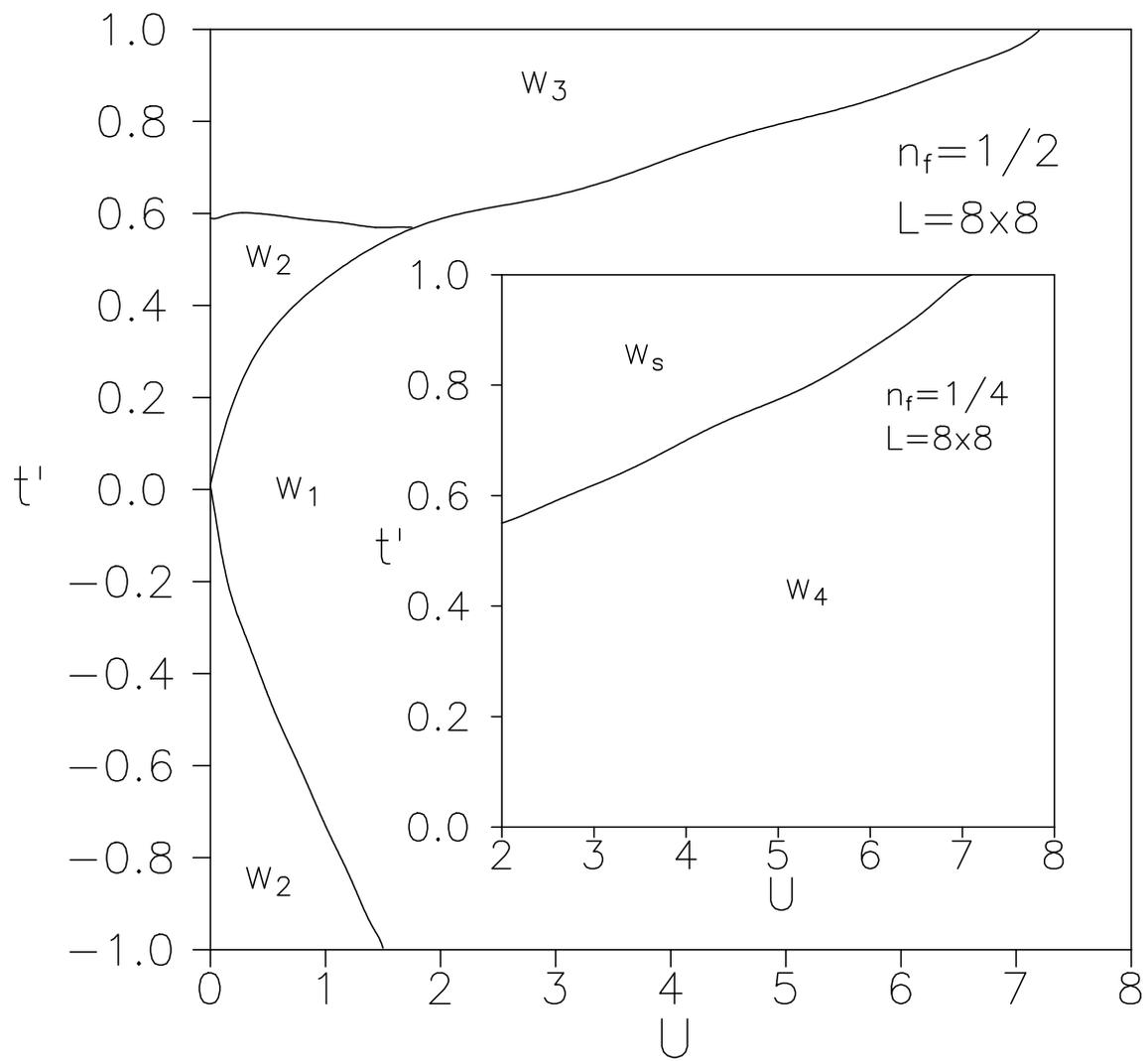
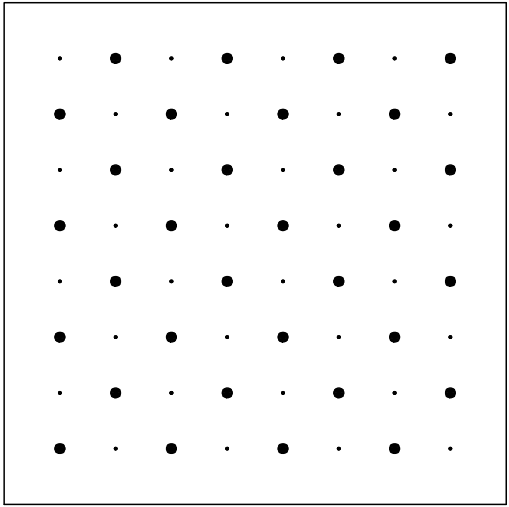
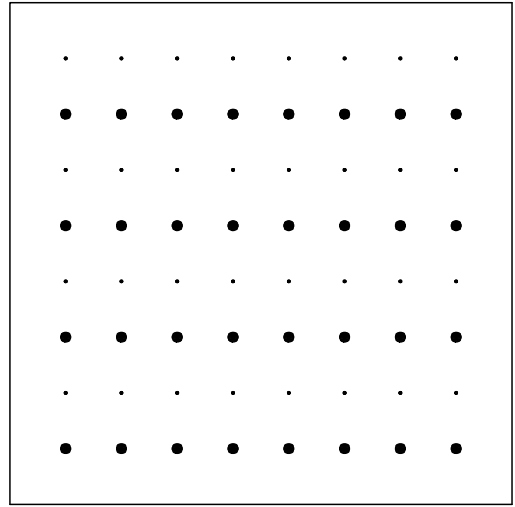
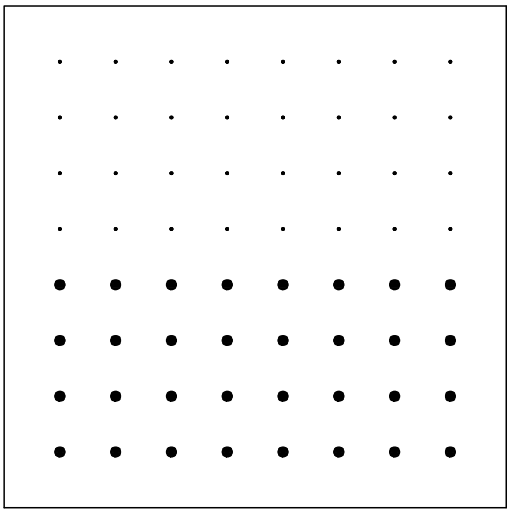
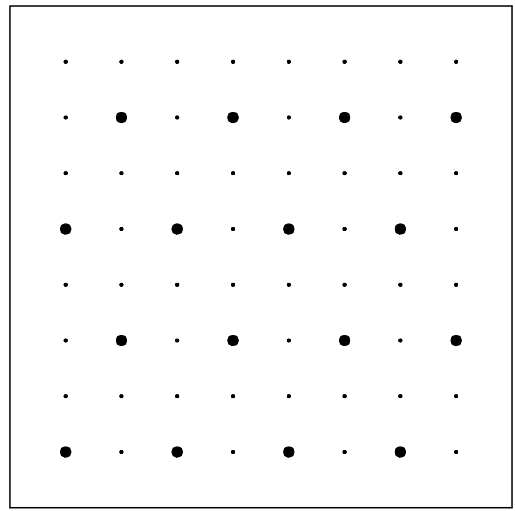
\end{document}